# Correction of Teledyne Acoustic Doppler Current Profiler (ADCP) Bottom-Track Range Measurements for Instrument Pitch and Roll

Rebecca A. Woodgate[1] and Alexander E. Holroyd[2]
[1]Applied Physics Laboratory, University of Washington, Seattle, WA. USA;
[2]Microsoft Research, Redmond, WA, USA
Corresponding author: Rebecca Woodgate, woodgate@apl.washington.edu

**Summary:** The Workhorse Acoustic Doppler Current Profiler (ADCP) manufactured by Teledyne RD Instruments (RDI) uses a "Bottom-Tracking" algorithm to yield data intended to represent distance to the bottom. However, current RDI software processing does not take into account the pitch and roll of the instrument. This technical note outlines post-deployment computations required to correct the reported Bottom-Track Ranges for instrument pitch and roll in the scenario where the ADCP is upward-looking with Bottom-Tracking being used to estimate distance to the surface.

## 1. Background

In "Bottom-Track" mode, Teledyne RD Instruments (RDI) Acoustic Doppler Current Profilers (ADCPs) report four "Bottom-Track Range" measurements of distance to the surface, one from each of the four beams of the instrument. From the ADCP Practical Primer (Teledyne RD Instruments, 2011), page 19:

> "*The bottom-track coordinate transformation is identical to the one used for the water profile. Bottom-track output also includes the vertical component of the distance, along each beam, to the bottom.*"

However, investigation of moored upward-looking ADCP data and subsequent discussions with Teledyne indicate that the data reported as Bottom-Track Range are corrected only for nominal beam angle, not for actual pitch and roll of the ADCP. Quoting email from Gregory Rivalan, Teledyne-RDI, 26th Sept 2011:

> "*I was able to confirm with our Firmware Engineers that the ranges are vertical but Tilts are not applied. That is we apply the cos(beam angle) to the slant range [WHDVLs Beam angle=30deg and WHADCPs=20deg] but the tilts are not applied.*"

This is confirmed by examining data (converted to ASCII by RDI's BBLIST software) from a 300kHz upward-looking ADCP moored at ~ 37m depth from summer 2007 to summer 2008 in the Bering Strait at site A4-07 (latitude: 65º 44.77' N, long: 168º 15.77' W), Figure 1.

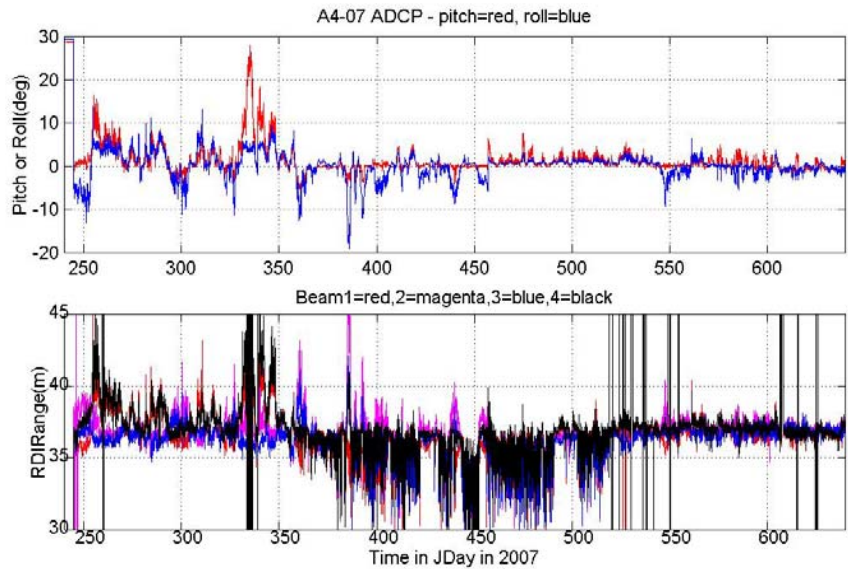

*Figure 1:* Upper panel - instrument pitch and roll. Lower panel - the four Bottom-Track (BT) Ranges. BT Ranges shorter than the instrument depth (~ 37m) between Julian Day ~ 380-490 indicate the presence of ice. BT Ranges agree only when pitch/roll are small. At larger pitch/roll angles, there are significant discrepancies in BT Ranges reported from the four beams. Changes are consistent with slant range changes resulting from instrument tilting. The mode of the BT Ranges is around 37m, confirming that the reported BT Range is not the slant range along the beam (37m/cos(20deg)= 39m), but has been corrected nominally by the instrument beam angle as per the email from RDI.

## 2. Formula for correction of RDI Bottom-Track Range for pitch and roll

To make the correction, it is necessary to obtain from pitch and roll measurements the inclination of each beam of the ADCP to the vertical. From Teledyne 2011, page 24, the pitch and roll inclinometers "measure tilt relative to earth gravity, but cannot differentiate the acceleration of gravity from accelerations caused, for example, by surface waves". Also from Teledyne 2011, page 18, looking down on the ADCP head, the beams (B1, B2, B3, B4) are numbered as follows:

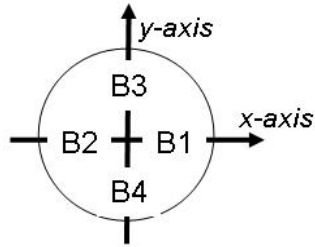

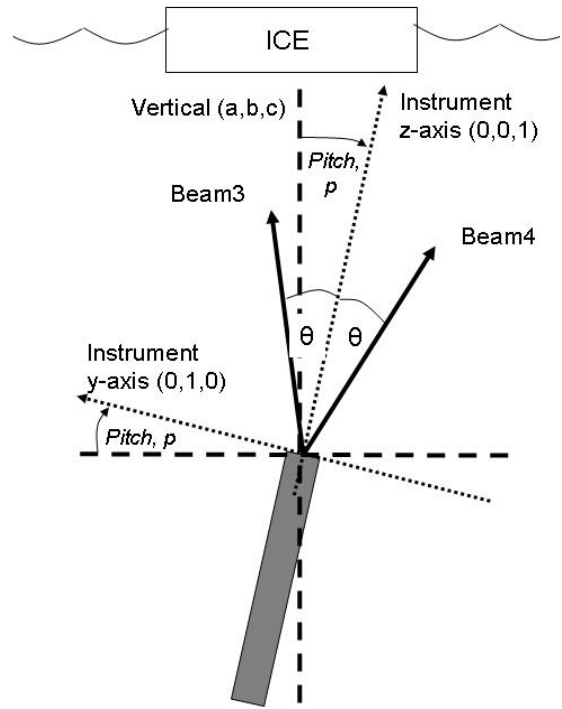

with B3 defined as the forward direction.

The pitch is the angle of the axis through B4 and B3 to the horizontal, and by inspection positive pitch means the head of B3 is higher than the head of B4 when the ADCP is upward-looking (see right for schematic).

Similarly, the roll is the angle of the axis through B2 and B1 to the horizontal and by inspection positive roll means the head of B2 is higher than the head of B1 when the ADCP is upward-looking.

One natural approach is to apply two successive rotations (corresponding to pitch and roll) to the beam directions. However, this is problematic owing to the non-commutativity of rotations in three dimensions. Instead we proceed as follows.

It is convenient to consider the problem in the coordinate frame of the instrument. In this frame, the instrument may be described as a vector:

  Instrument = ( 0, 0, 1 ),

where the 3$^{rd}$ (z) component is along the axis of the instrument. In the upward-looking scenario under consideration here, we pick arbitrarily the x-axis to be towards B1 (and away from B2), and the y-axis to be towards B3 (and away from B4), and the z-axis to be along the instrument in the direction of the outgoing beams.

The beams are at an angle $\theta$ to the axis of the instrument ($\theta$ being 20deg for a Workhorse ADCP). Thus, in this instrument coordinate frame, the beams can be represented as unit-length vectors:

$$
\begin{aligned}
\text{Beam1} &= (\ +\sin\theta,\quad\ \ 0,\ +\cos\theta\ ) \\
\text{Beam2} &= (\ -\sin\theta,\quad\ \ 0,\ +\cos\theta\ ) \\
\text{Beam3} &= (\quad\ \ 0,\ +\sin\theta,\ +\cos\theta\ ) \\
\text{Beam4} &= (\quad\ \ 0,\ -\sin\theta,\ +\cos\theta\ ).
\end{aligned}
\qquad(1)
$$

We now seek the true vertical, which we denote by vector V = ( a, b, c ) in the instrument frame, where a, b, and c are unknown. By requiring the length of V to be unity without loss of generality:

$$a^2 + b^2 + c^2 = 1 \, , \qquad \text{thus} \quad c = \sqrt{1 - a^2 - b^2} \, . \tag{2}$$

The pitch, p, of the instrument is the angle between the instrument y-axis ( 0, 1, 0 ) and the horizontal. Thus $\pi/2 - p$ is the angle between our instrument y-axis and the true vertical ( a, b, c ).

Taking the dot product of our instrument y-axis and the true vertical yields:

$$( 0, 1, 0 ) \cdot ( a, b, c ) = b = \cos(\pi/2 - p) = \sin p \, . \tag{3}$$

Similarly for the roll, r, which is relative to our instrument x-axis:

$$( 1, 0, 0 ) \cdot ( a, b, c ) = a = \cos(\pi/2 + r) = -\sin r \, . \tag{4}$$

Note the sign conventions here match the RDI-ADCP sign conventions for pitch and roll.

We seek the angle α between each beam and the true vertical, the cosine of which is obtained from the dot product of the Beam (Equation 1) with the true vertical V. For example, for Beam 1, using Equations 2, 3 and 4 for c, b and a,

$$\begin{aligned}
\cos \alpha_1 &= (\text{Beam 1}) \cdot (V) = ( +\sin\theta, \, 0, \, +\cos\theta ) \cdot ( a, b, c ) \\
&= a \sin\theta + c \cos\theta \\
&= -\sin r \, \sin\theta + \cos\theta \sqrt{1 - \sin^2 r - \sin^2 p} \, .
\end{aligned}$$

This process, repeated for all beams, yields the following expressions for the angles of the beams to the vertical:

$$\begin{aligned}
\text{Beam 1:} \quad \cos \alpha_1 &= -\sin r \, \sin\theta + \cos\theta \sqrt{1 - \sin^2 r - \sin^2 p} \, , \\
\text{Beam 2:} \quad \cos \alpha_2 &= +\sin r \, \sin\theta + \cos\theta \sqrt{1 - \sin^2 r - \sin^2 p} \, , \\
\text{Beam 3:} \quad \cos \alpha_3 &= +\sin p \, \sin\theta + \cos\theta \sqrt{1 - \sin^2 r - \sin^2 p} \, , \\
\text{Beam 4:} \quad \cos \alpha_4 &= -\sin p \, \sin\theta + \cos\theta \sqrt{1 - \sin^2 r - \sin^2 p} \, .
\end{aligned} \tag{5}$$

The above solution is exact, whereas Shcherbina *et al.*, 2005 present an approximate solution to the same problem in which interactions between pitch and roll are not taken into account.

The corrected vertical ranges from the ADCP to the surface (Vrange_1-4) are now obtained from the reported RDI Bottom-Track Ranges (RDIRange_1-4) as follows:

$$\text{Vrange\_i} = \cos \alpha_i \, \text{RDIRange\_i} \, / \cos\theta \, , \quad \text{for i = 1, 2, 3, 4} \tag{6}$$

where θ is 20deg for a 300 kHz Workhorse. (The factor cos θ in the denominator compensates for the scaling applied to the data by the RDI software.)

## 3. Application of correction to upward-looking ADCP data from mooring site A4-07

To assess the importance of this correction, we consider the case of the ADCP at A4-07 introduced above, which was moored upward-looking ~ 37m below the surface. From Teledyne, 2011, page 44, Bottom-Tracking "depth resolution is approximately 0.1m". From RD Instruments, 2001, page 27, Bottom-Track data using ADCP option BM4 are claimed to be reported in increments of 1% of range (for 40m, thus 0.4m), however the ASCII data obtained via BBLIST are reported to greater precision than this. The tilt sensors on the Workhorse have a range of ± 15deg, with an uncertainty of ± 2% (RD Instruments, 1998, page A5), presumably 2% of 30deg, i.e., 0.6deg.

Simple geometry suggests that neglecting tilts (t) of 5, 10 and 15 degrees in an ADCP moored 40m below the surface results in a range errors of [ depth x cos(θ+t) / cos θ ] ~ 1.4m, 3.4m, 5.9m. Figure 2 (below) shows the distribution of the RDI Bottom-Track Ranges from the A4-07 ADCP as a function of roll (top line) and pitch (bottom line) for the different beams. Blue dots mark all data. Cyan and red dots indicate times where the other tilt parameter is less than 0.5deg or 0.1deg respectively, i.e., times where change in range are due to tilt in one component only. Note that variation in roll at zero pitch changes Beam 1 and 2 ranges, but not Beam 3 and 4 ranges. Similarly, variation in pitch at zero roll affects only Beams 3 and 4. Again the short ranges (around pitch/roll of zero) likely relate to ice keels.

*Figure 2*: *See main text above.*

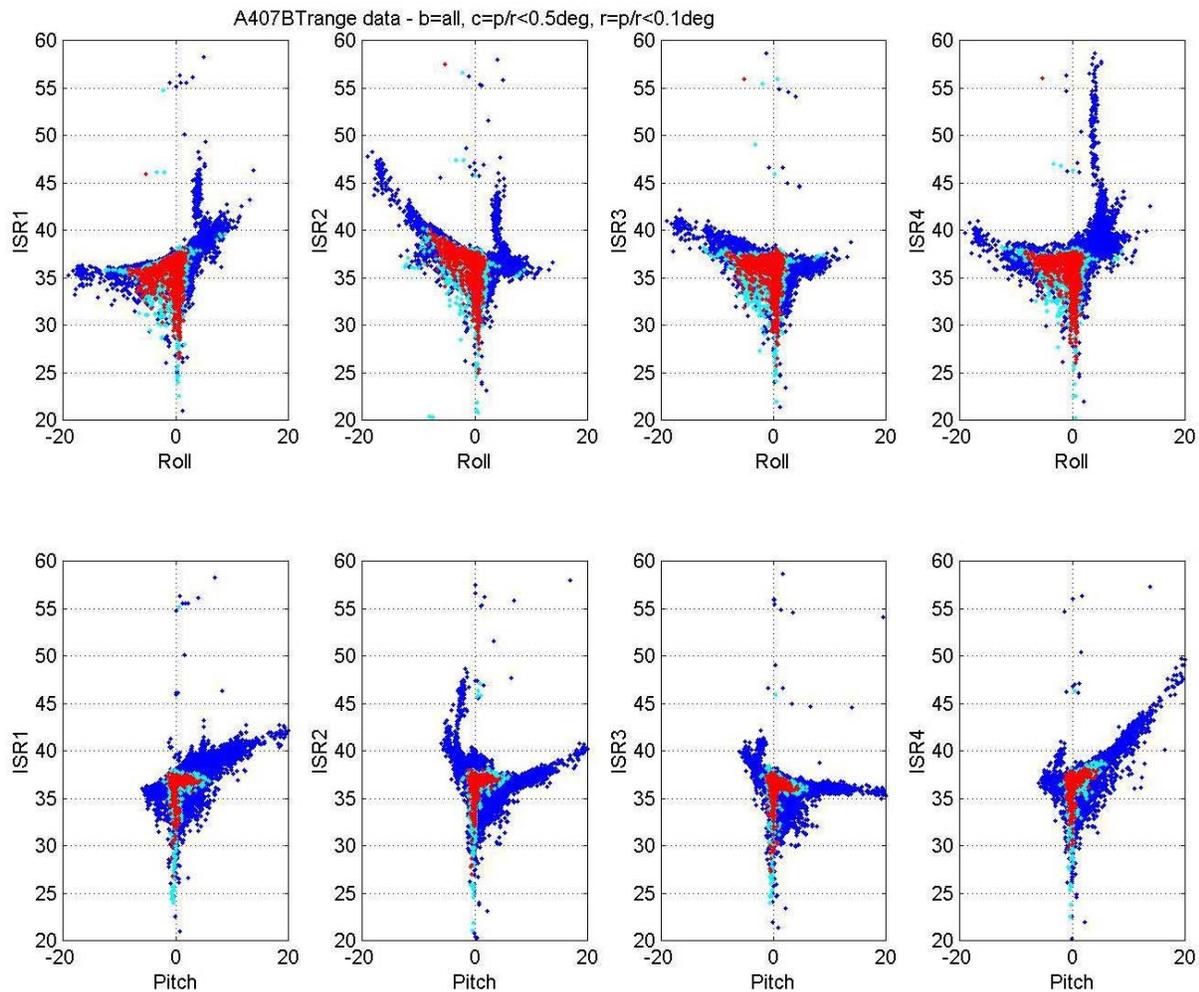

Consider now (Figure 3) the same data corrected for pitch and roll using Equations 5 and 6 above. In Figure 3, black dots mark uncorrected data, red marks corrected data with tilts ≤15deg and cyan marks data with tilts >15deg, i.e., outside the published range of the tilt sensors. The corrected ranges are now almost independent of roll (top row). The dependency on pitch is significantly reduced, but is still present at large pitches, and is comparable to the quoted uncertainty of 1% of range, i.e. in this case ~ 0.4m (marked in Figure 3, lower panels).

*Figure 3:* See main text. Lower 2 rows are expanded from upper rows, and also mark 0.4m range error.

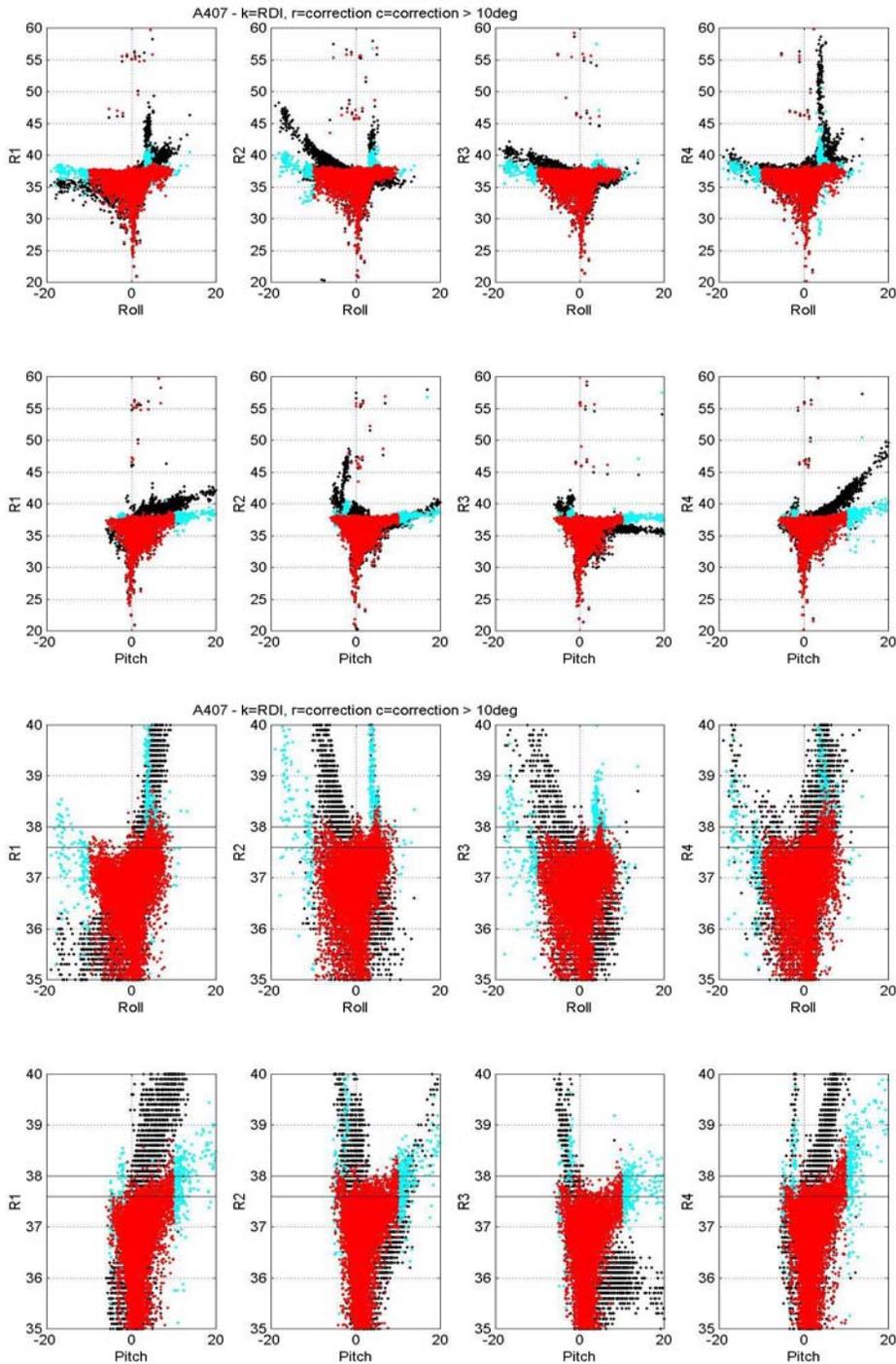

There are two obvious suspects for this remaining dependency on pitch. The first is inaccuracy in the pitch and roll sensors. From the manufacturer's specifications, errors in pitch and roll may be 0.6deg. This magnitude of error results in (not shown) range errors of ~ 0.15m. The second suspect is error in the beam headings. Again, an error of 0.6deg results in range errors of ~ 0.15m. Combined, these two factors result in errors ~ 0.3m, with this error extending at high pitch angles to almost 0.6m.

*Estimating Pitch and Roll Corrections:* We next examine if constant offsets in pitch and roll are a significant source of error. Such offsets can be estimated from the data since geometrically the pitch and roll of the instrument may be calculated from the four beam ranges if the surface is flat. We consider first the case where roll is zero, and thus the instrument inclination is the pitch, p. The Beams 3 and 4, angled at $\theta$ to the instrument, are then at $\theta+p$ to the vertical (Beam 4) and at $\theta-p$ to the vertical (Beam 3). If D is the true depth of the instrument below the surface, then:

$$\cos(\theta + p) = D / b4 \quad \text{(where b4 = slant range of Beam 4)},$$
$$\cos(\theta - p) = D / b3 \quad \text{(where b3 = slant range of Beam 3)}. \tag{7}$$

Dividing to remove D and rearranging yields:

$$\tan p = \frac{1}{\tan\theta} \frac{(b4 - b3)}{(b4 + b3)} \quad \text{when roll, r, is zero.}$$

Similarly, at times when pitch is zero, a similar equation can be derived for the roll, r, viz:

$$\tan r = \frac{1}{\tan\theta} \frac{(b1 - b2)}{(b1 + b2)} \quad \text{when pitch, p, is zero.}$$

The approximation here is as per Shcherbina *et al.,* 2005, Equation 16, although beware an ambiguity in their brackets. The full problem for non-zero pitch and roll is not readily solved analytically and is not pursued here.

Figure 4 (below) shows the discrepancy between these estimates of pitch and roll and the measured pitch and roll, plotted against the tilt components. Grey indicates all data. Red indicates points where the non-adjusted tilt component is < 0.2deg, blue is the subset of those points at times believed to be ice free. Lower panels are enlargements of the 0-5deg range. Significant scatter is evident in these plots, but to seek a constant offset, we consider the mean of blue points in this reduced range. This yields offsets of:
        for pitch = - 0.17deg (i.e., RDI pitch estimate is less than calculated pitch), and
        for roll   = + 0.40deg (i.e., RDI roll estimate is greater than calculated roll).
Both numbers are within stated instrument uncertainties.

Including these estimates as constant offsets to the range estimates as per Figure 3 does not significantly reduce the dependency on pitch, suggesting that pitch dependent errors in the sensors may be significant. However, this possibility is not pursued here, since the majority of the data are at small tilts, specifically:

- 91% of all pitch readings are less than 5deg,
- 88% of all roll readings are less than 5deg,

- 96% of all pitch readings are less than 10deg,
- 97% of all pitch readings are less than 10deg.

*Figure 4:* See main text above.

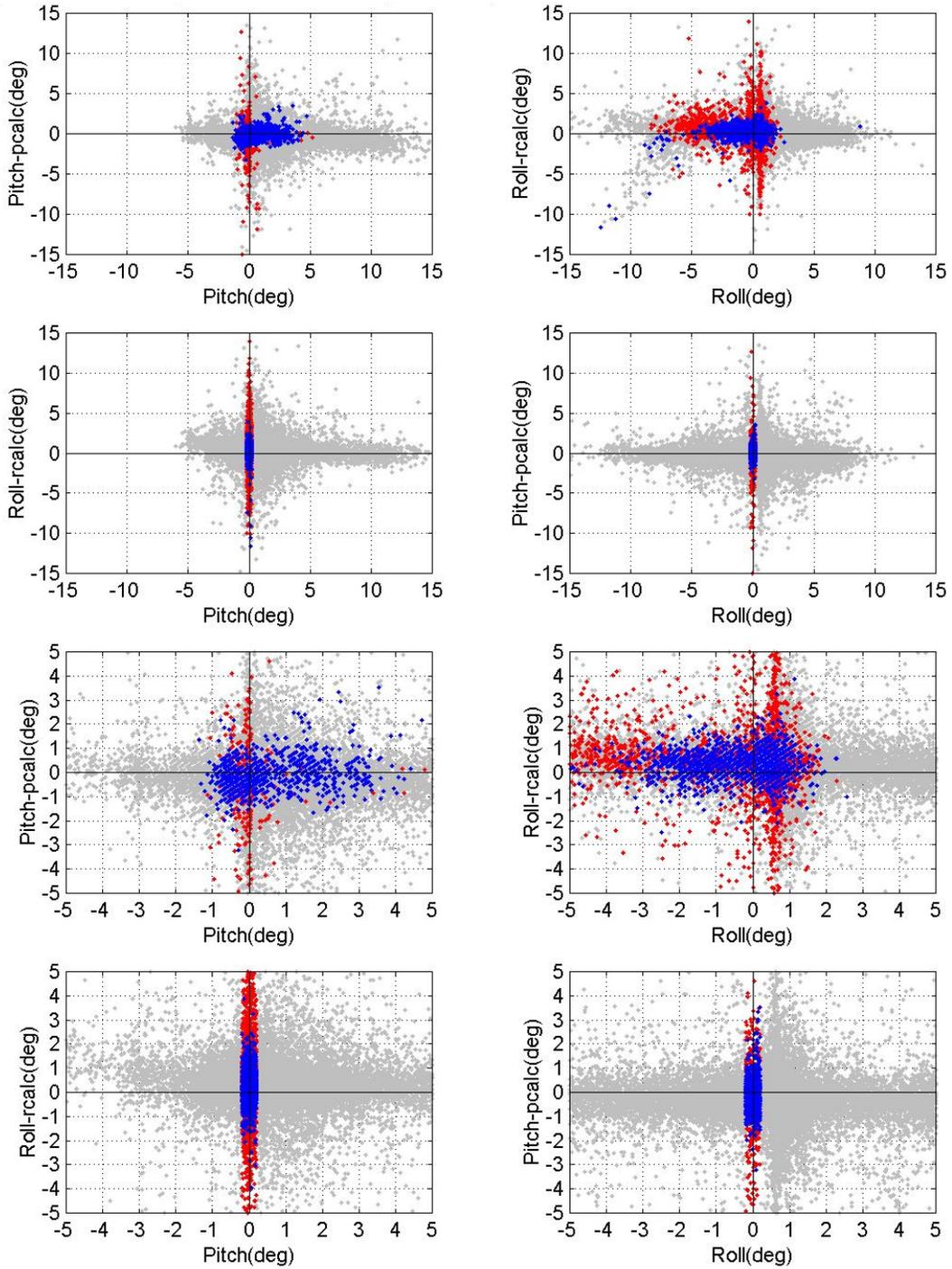

***Estimating Beam Angle Corrections:*** Another source of error may be misalignments in the beam heads. This is investigated by assuming pitch and roll are correct and solving Equation 7 for θ rather than p (and similarly for r). Histograms of the solutions for θ (Figure 5, right) suggest θ is perhaps 1-2deg greater than the nominal 20deg.

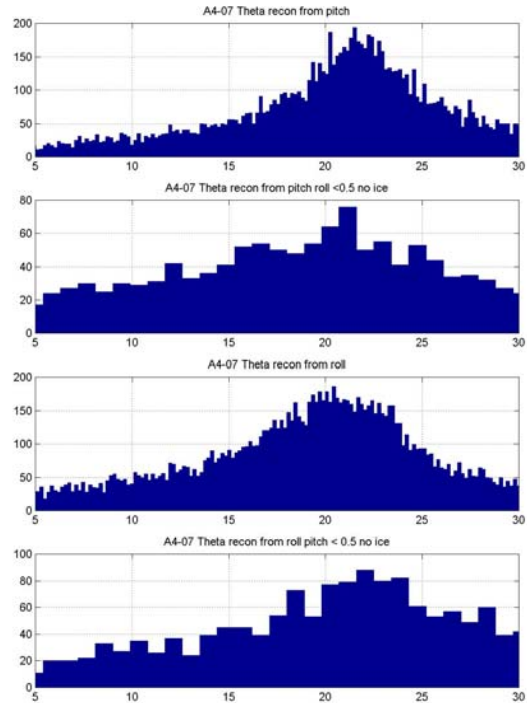

***Figure 5:*** *Histograms of the solution for θ calculated as described in text from pitch (top 2 panels) and roll (bottom 2 panels). The 1$^{st}$ and 3$^{rd}$ panels are for the entire record. The 2$^{nd}$ and 4$^{th}$ panels are for ice-free times when the other tilt parameter is <0.5deg.*

To date, it is not possible to confirm this offset with calibration data from the instrument. The instrument's transformation matrix (obtained via the PS3 test) is identical to the ideal 20deg solution (Teledyne, 2010, page 11) and is accompanied by the information that "beam angle corrections are loaded", implying that beam head errors are negligible. In any case, applying these corrections to the data still does not lead to a noticeable improvement in the pitch dependency (Figure 6, below).

***Figure 6:*** *Ranges from different beams (columns) plotted against roll (top row) and pitch (bottom row). Grey marks uncorrected data. Blue indicates geometric correction for pitch and roll assuming no instrument errors. Red is further corrected for pitch, roll and θ offsets as discussed above. Horizontal lines mark 0.4m ranges from 38m.*

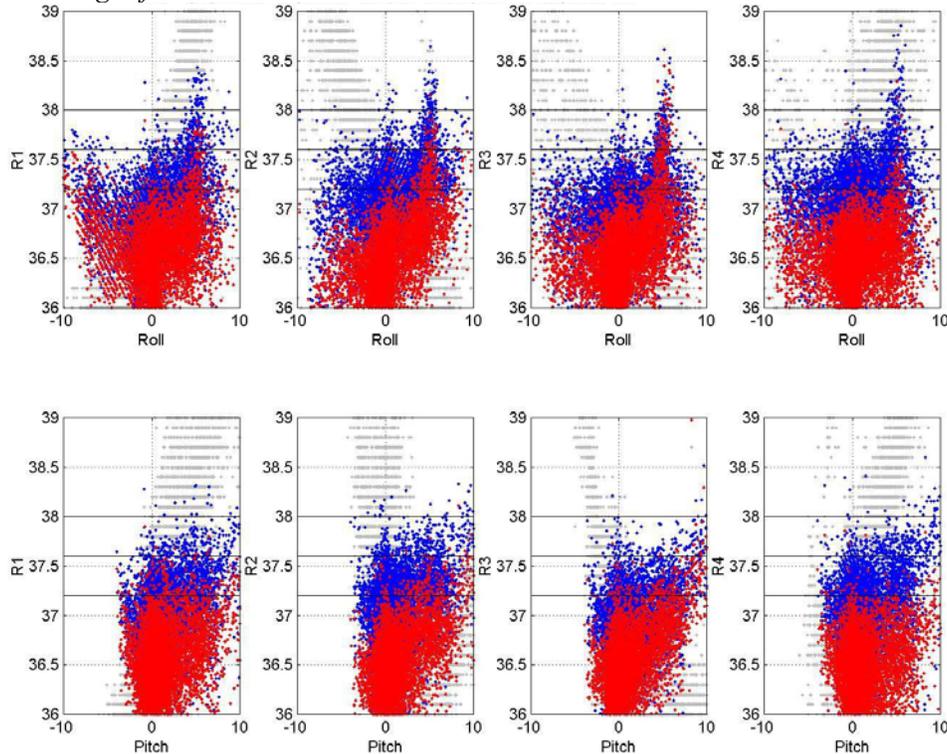

## 4. Conclusions

We conclude that it is essential (and straightforward) to correct geometrically for tilt and roll for ADCP Bottom-Track Ranges, as per Equations 5 and 6. A summary figure of the corrections is given below (Figure 7) and shows significant improvement in the ranges, manifested by greater agreement in the ranges from different beams.

In this example, the data once corrected still show some (greatly reduced) dependency on pitch and roll which is likely due to errors in the tilt sensors and/or in the deviation of the beams from the nominal 20deg orientation. While simple offsets may be calculated, they do not appear to solve this problem.

The analysis to date suggests that the quoted errors of 0.4m in the Bottom-Track Range are a reasonable quantification of errors only once the pitch and roll of the instrument have been taken into account.

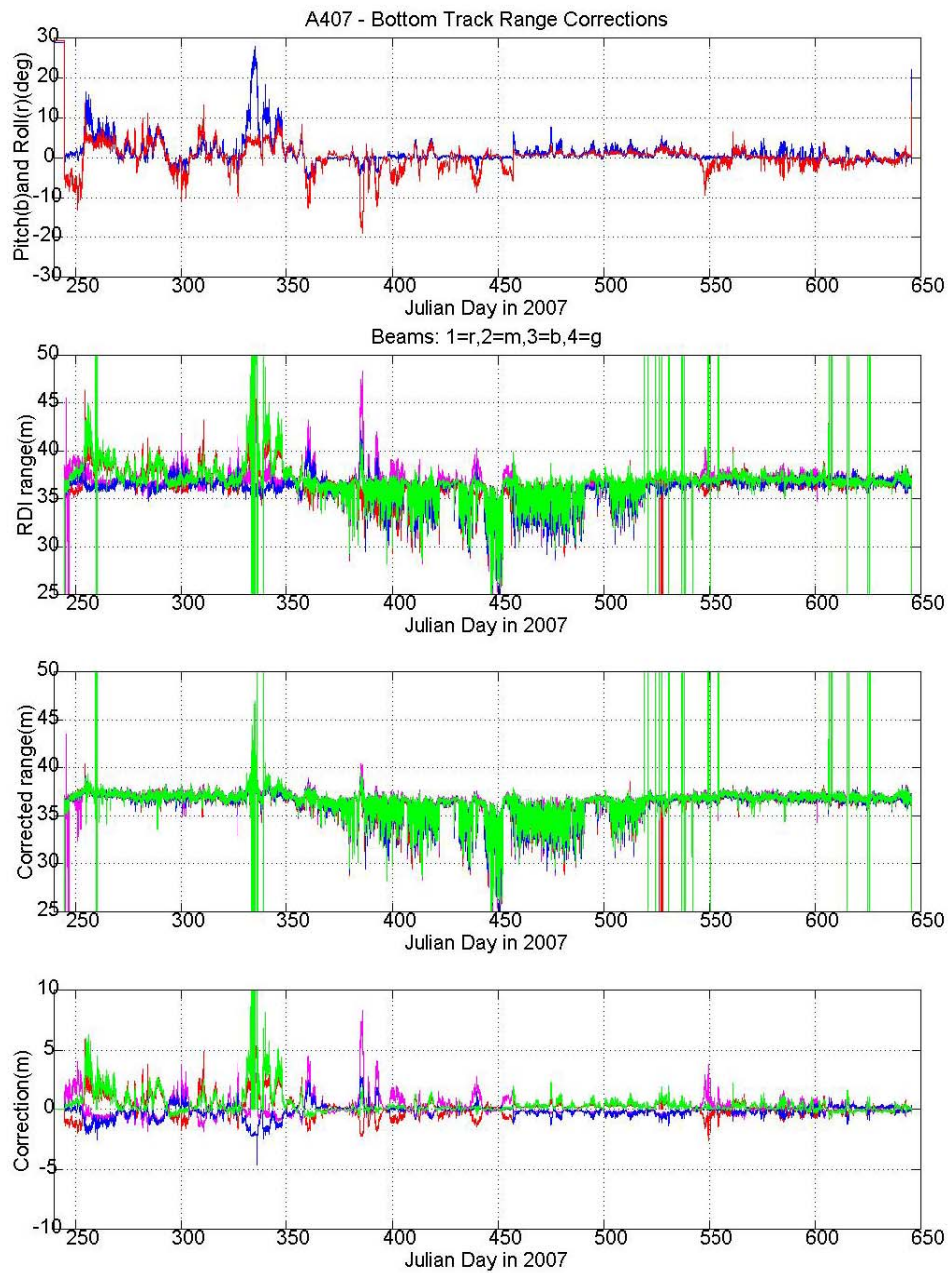

*Figure 7:*
*Top panel – pitch (blue) and roll (red) reported by the RDI in degrees.*

*Second panel – Bottom-Track Ranges reported directly from the instrument (i.e., uncorrected for pitch and roll). Colors indicate different beams.*

*Third panel – ranges corrected for pitch and roll as per Equations 5 and 6. Colors indicate different beams.*

*Fourth panel – difference between the RDI reported Range to surface and the range to surface calculated taking into account the pitch and roll. Colors indicate different beams.*